\documentclass[amsmath,amssymb,nofootinbib,showkeys]{revtex4}
\usepackage{graphicx}
\usepackage{dcolumn}
\usepackage{color}
\usepackage{bm}
\usepackage{amsmath,amssymb,graphicx}
\usepackage{epsfig}
\usepackage{enumerate}
\usepackage{array}
\usepackage{multirow}

\newtheorem{theorem}{Theorem}

\newtheorem{prop}{Proposition}

\begin{document}
\title{Jacobi last multiplier and two-dimensional superintegrable oscillators}

\author{Akash Sinha\footnote{E-mail: s23ph09005@iitbbs.ac.in} and Aritra Ghosh\footnote{E-mail: ag34@iitbbs.ac.in}}

\vspace{2mm}

\affiliation{School of Basic Sciences, Indian Institute of Technology Bhubaneswar, Jatni, Khurda, Odisha 752050, India}

\vskip-2.8cm
\date{\today}
\vskip-0.9cm


\begin{abstract}
In this paper, we examine the role of the Jacobi last multiplier in the context of two-dimensional oscillators. We first consider two-dimensional unit-mass oscillators admitting a separable Hamiltonian description, i.e., \(H = H_1 + H_2\), where \(H_1\) and \(H_2\) are the Hamiltonians of two one-dimensional unit-mass oscillators; it is shown that there exists a third functionally-independent first integral \(\Theta\), thereby ensuring superintegrablility. Various examples are explicitly worked out. We then consider position-dependent-mass oscillators and the Bateman pair, where the latter consists of a pair of dissipative linear oscillators. Quite remarkably, the Bateman pair is found to be superintegrable, despite admitting a Hamiltonian which cannot be separated into those of two isolated (non-interacting) one-dimensional oscillators. 
 \end{abstract}


\maketitle

\section{Introduction}\label{introsec}
In the context of Hamiltonian systems where the phase space is of dimension \(m = 2n\), there is the notion of Liouville-Arnold integrability \cite{arnoldcm} which requires only \(n\) functionally-independent first integrals but with mutually-commuting Poisson brackets, i.e., the set \(\{F_j\}\), such that \(\{F_j, F_k\}_{\rm P.B.}=0\), \(\forall\) \(j,k \in \{1,2,\cdots,n\}\). Here, the subscript `P.B.' indicates the (canonical) Poisson bracket. However, often enough, integrable Hamiltonian systems admit more than \(n\) functionally-independent first integrals; well-known examples include the \(n\)-dimensional linear oscillator and the Kepler problem, with symmetry groups \(SU(n)\) and \(SO(4)\), respectively \cite{goldsteincm,ANISO,ANISO1}. Such systems are called superintegrable \cite{evans,marquette}. A maximally-superintegrable system is one that admits the maximum number of independent conserved quantities, which, on a phase space of dimension \(2n\), is \(2n-1\). Thus, the \(n\)-dimensional linear oscillator and the Kepler problem are maximally superintegrable. \\

An interesting class of superintegrable Hamiltonian systems are two-dimensional oscillators with the configuration space \(Q\) being a subset of \(\mathbb{R}^2\). The phase space is therefore, a submanifold of \(T^*\mathbb{R}^2\) and admits a canonical symplectic form, \(\omega_{\rm S} = \sum_{j=1}^{2} dq_j \wedge dp_j = dq_1 \wedge dp_1 + dq_2 \wedge dp_2\) in Darboux coordinates; notice that \((q_1,q_2)\) are coordinates in \(Q \subset \mathbb{R}^2\), while \((p_1,p_2)\) are the corresponding `induced' fiber coordinates on \(T^*Q \subset T^*\mathbb{R}^2\). Given a Hamiltonian function \(H \in C^\infty (T^*Q, \mathbb{R})\), the Hamiltonian vector field \(X_H\), whose integral curves are the solutions of the Hamilton's equations, is determined by the following intrinsic condition \cite{arnoldcm}:
\begin{equation}\label{intrinsic}
\iota_{X_H} \omega_{\rm S} = dH. 
\end{equation}
In the present case, the Hamilton's equations are
\begin{equation}\label{eqnofmotion}
\dot{q}_1 = \frac{\partial H}{\partial p_1}, \hspace{3mm} \dot{q}_2 = \frac{\partial H}{\partial p_2}, \hspace{3mm} \dot{p}_1 = -\frac{\partial H}{\partial q_1}, \hspace{3mm} \dot{p}_2 = -\frac{\partial H}{\partial q_2},
\end{equation} where \((q_1,q_2,p_1,p_2)\) are just the Cartesian coordinates in some region of \(T^* \mathbb{R}^2 \approxeq \mathbb{R}^4\). The phase space admits a Poisson structure, with the Poisson bracket being defined as \(\{G_1,G_2\}_{\rm P.B.} = \omega_{\rm S}(X_{G_1},X_{G_2})\), for any \(G_1, G_2 \in C^\infty(T^*Q,\mathbb{R})\). In coordinates \((q_1,q_2,p_1,p_2)\), it reads
\begin{equation}
\{G_1,G_2\}_{\rm P.B.} = \frac{\partial G_1}{\partial q_1}\frac{\partial G_2}{\partial p_1} - \frac{\partial G_1}{\partial p_1}\frac{\partial G_2}{\partial q_1} + \frac{\partial G_1}{\partial q_2}\frac{\partial G_2}{\partial p_2} - \frac{\partial G_1}{\partial p_2}\frac{\partial G_2}{\partial q_2} .
\end{equation} 
The subject of two-dimensional superintegrable oscillators is currently being worked on, with various recent developments \cite{2d0,2d1,2d2,2d3,2d4,2d5,2d6,2d7,2d8,2d9,2d10,2d11,2d12,2d13,2d14,2d15}. \\

It may be remarked that for two-dimensional oscillators, the notion of superintegrability coincides with that of maximal superintegrability, because the phase space is four-dimensional and on it, one can have only a maximum of three independent conserved quantities. Such systems are of physical interest \cite{superphys}, for a classical system which is (maximally) superintegrable has its trajectories confined to one-dimensional orbits which necessarily close if they are confined. Moreover, quantum mechanically, maximal superintegrability is associated with quasi-exact solvability and often the energy levels can be calculated algebraically, say, in the case of the \(n\)-dimensional linear oscillator. \\

In this paper, we shall be primarily interested in systems which are just two copies of integrable one-dimensional oscillators, i.e., the Hamiltonian is \(H = H_1 + H_2\), where \(H_1 = H_1(q_1,p_1)\) and \(H_2 = H_2 (q_2,p_2)\). This would invariably imply that one has two independent first integrals, \(F_1 = H_1\) and \(F_2 = H_2\) to start with, thereby ensuring integrability in the Liouville-Arnold sense. However, as we shall show, such systems are superintegrable and admit a third (independent) constant of motion, \(F_3 = \Theta\). For this, we compute a function known as the Jacobi last multiplier \cite{JLM1,JLM2,JLM3,JLM4,JLM5,JLM6,JLM7,JLM8}, which shall allow us to find the third integral of motion. Typically, the last multiplier as it appears in Whittaker's textbook \cite{JLM1} allows one to determine the `last' integral of motion on an \(m\)-dimensional phase space (\(m\) may or may not be even) when \((m-2)\) first integrals are already known. We apply the formalism to find the third integral of motion of some two-dimensional oscillators of the above-mentioned type, i.e., with \(H = H_1 + H_2\). In all such cases, we demonstrate that the conserved quantity \(\Theta\) is just the difference between the angle variables associated with the action variables \(I_1 \sim H_1\) and \(I_2 \sim H_2\), which is consistent with the analysis presented in \cite{2d6,2d7,2d8,2d9,2d10,2d11}. Following this, we consider the Bateman pair, which corresponds to a pair of (linear) dissipative oscillators such that one of them loses energy with a certain `loss' coefficient, while the other gains energy with a `gain' coefficient of the same magnitude. We show that although this system cannot simply be expressed as a sum of two non-interacting oscillators, the pair is superintegrable. \\

With this background, let us now present the organization of this paper. In the next section [Section (\ref{DSsec})], we briefly review the Jacobi last multiplier, and its use in determining the last integral of motion, to be denoted by \(\Theta\). In particular, in Section (\ref{DSsec1}), we focus on two-dimensional separable Hamiltonian systems representing unit-mass oscillators, and work out the special case of a two-dimensional (isotropic) linear oscillator to demonstrate the method. Then, in Section (\ref{consec}), we use the method of the last multiplier to determine \(\Theta\) for a few two-dimensional systems, establishing superintegrability of the systems. Following that, in Sections (\ref{PDMsec}) and (\ref{Bsec}), we discuss the superintegrability of position-dependent-mass oscillators and the Bateman pair, respectively. We end with some discussion in Section (\ref{dsec}).

\section{Jacobi last multiplier and first integrals}\label{DSsec}
Consider some autonomous dynamical system on \(\mathbb{R}^m\) which can be generically expressed as
\begin{equation}\label{genericdynamicalsystem}
\dot{x}_j = W_j (x_1,x_2, \cdots, x_m),
\end{equation} where \(W_j: \mathbb{R}^m  \rightarrow \mathbb{R}\) and \(j \in \{1,2, \cdots, m\}\). Now, let \((F_1,F_2,\cdots,F_k)\) be the set of first integrals of the dynamical system (\(k < m\)). For any open subset \(\Omega \subset \mathbb{R}^m\), we define the Jacobi last multiplier to be a function \(M: \mathbb{R}^m \rightarrow \mathbb{R}\), which is non-negative and defines an invariant measure \(\int_\Omega M d^m x\), i.e.,
\begin{equation}
\int_\Omega M d x_1 \wedge  d x_2 \wedge \cdots \wedge d x_m = \int_{\phi_t({\Omega})} M \frac{\partial(x_1,x_2,\cdots,x_k,x_{k+1},\cdots, x_m)}{\partial(F_1,F_2,\cdots,F_k, x_{k+1},\cdots, x_m)} d F_1 \wedge dF_2 \wedge  \cdots \wedge d F_k \wedge dx_{k+1} \wedge \cdots dx_m,
\end{equation} where \(\phi_t(\Omega)\) is the transformation of the region \(\Omega\) under the flow of the vector field whose components are \(\{W_j\}\). It should be noticed that it is necessary that \((F_1,F_2,\cdots,F_k)\) must be independent, implying that \(dF_1 \wedge dF_2 \wedge \cdots \wedge dF_k \neq 0\). The above-mentioned invariance condition leads to the equation 
\begin{equation}
\frac{d}{dt} \Bigg( M \frac{\partial(x_1,x_2,\cdots,x_k,x_{k+1},\cdots, x_m)}{\partial(F_1,F_2,\cdots,F_k, x_{k+1},\cdots, x_m)}  \Bigg) = 0,
\end{equation} which, using chain rule gives \cite{JLM7}
\begin{equation}
\frac{dM}{dt}  \frac{\partial(x_1,x_2,\cdots,x_k,x_{k+1},\cdots, x_m)}{\partial(F_1,F_2,\cdots,F_k, x_{k+1},\cdots, x_m)}  + M \sum_{j=1}^m \frac{\partial W_j}{\partial x_j}  \frac{\partial(x_1,x_2,\cdots,x_k,x_{k+1},\cdots, x_m)}{\partial(F_1,F_2,\cdots,F_k, x_{k+1},\cdots, x_m)}= 0.
\end{equation} This is just equivalent to 
\begin{equation}\label{meqn}
\frac{d}{dt} \ln M + \sum_{j=1}^m \frac{\partial W_j}{\partial x_j} = 0,
\end{equation}
which tells us that the last multiplier is related to the divergence of the dynamical vector field. In fact, if the last multiplier is known along with \((m-2)\) first integrals, then it lets one determine the \((m-1)\)th or equivalently, the `last' first integral. Consider
\begin{equation}\label{fkrel}
F_k(x_1,x_2,\cdots,x_m) = c_k, \hspace{6mm} k \in \{1,2,\cdots,m-2\}.
\end{equation} The real constants \(\{c_k\}\) are just the on-shell values of the first integrals \(\{F_k\}\). We now perform a change of variables as
\begin{equation}
(x_1,x_2, \cdots, x_m) \mapsto (c_1,c_2,\cdots, c_k, \xi_1,\xi_2),
\end{equation} making use of the relations given in Eq. (\ref{fkrel}). Using the constants of motion, we may convert the dynamical system given by Eq. (\ref{genericdynamicalsystem}) into a planar one as
\begin{equation}\label{planardynamics}
\frac{d\xi_1}{dt} = \overline{X}_1 (\xi_1,\xi_2, \{c_k\}), \hspace{6mm} \frac{d\xi_2}{dt} = \overline{X}_2 (\xi_1,\xi_2, \{c_k\}),
\end{equation} where \(\overline{X}_1, \overline{X}_2: \mathbb{R}^2 \rightarrow \mathbb{R}\), are functions of the variables \((\xi_1,\xi_2) = (x_{m-1},x_m)\), obtained from Eq. (\ref{genericdynamicalsystem}) via elimination of \((m-2)\) variables. Then, upon defining
\begin{equation}
\Delta:=\frac{\partial(F_1,F_2,\cdots,F_{m-2}, x_{m-1}, x_m)}{\partial(x_1,x_2,\cdots,x_{m-2}, x_{m-1}, x_m)} = \frac{\partial(F_1,F_2,\cdots,F_{m-2})}{\partial(x_1,x_2,\cdots,x_{m-2})},
\end{equation}
 it is possible to show after some manipulations that \cite{JLM1,JLM7}
\begin{equation}\label{thetadef}
\Theta := \int \frac{\overline{M}}{\overline{\Delta}} (\overline{X}_1 d\xi_2 - \overline{X}_2 d\xi_1)
\end{equation} is a conserved quantity of the dynamical system given in Eq. (\ref{planardynamics}). Here, the `overline' denotes that \(\overline{M}\), \(\overline{\Delta}\), \(\overline{X}_1\), and \(\overline{X}_2\) are being considered after the \((m-2)\) variables have been eliminated. It should be emphasized that the coordinates \(\{x_1,x_2,\cdots,x_{m-2}\}\) should be chosen in such a way that \(\Delta \neq 0\).

\subsection{Two-dimensional separable Hamiltonian systems}\label{DSsec1}
Let us consider a two-dimensional standard Hamiltonian system, i.e., a system for which the Hamiltonian can be expressed as the sum of kinetic and potential energies. We further resort to the cases where a Cartesian (flat) coordinate system is involved, and where the Hamiltonian is \(H = H_1 + H_2\), with
\begin{equation}\label{H1H2}
H_1 = \frac{p_1^2}{2} + V_1(q_1), \hspace{6mm} H_2 = \frac{p_2^2}{2} + V_2(q_2).
\end{equation} Here, \(V_1 \in C^\infty(\Gamma_1,\mathbb{R})\) and \(V_2 \in C^\infty(\Gamma_2,\mathbb{R})\), where \(\Gamma_1, \Gamma_2 \subset \mathbb{R}\). Hamiltonian vector fields are volume preserving, as can be seen by constructing the Lie derivative, \(\pounds_{X_H} (\omega_{\rm S}) = \iota_{X_H} (d\omega_{\rm S}) + d(\iota_{X_H} \omega_{\rm S}) = 0\), where the first term vanishes because \(\omega_{\rm S}\) is closed, while the second term vanishes upon using Eq. (\ref{intrinsic}) and noticing that \(d^2 = 0\). Therefore, due to this volume-preserving nature of Hamiltonian vector fields, Eq. (\ref{meqn}) implies that \(M\) is a constant real number, which we take to be equal to one without loss of generality. \\

Now, for the system given by \(H = H_1 + H_2\), where \(H_{1,2}\) are of the form given in Eq. (\ref{H1H2}), the phase space is of real-dimension four, while we immediately have two functionally-independent first integrals, \(H_1\) and \(H_2\). This ensures that such systems are integrable. However, following the preceding discussion on the last multiplier, it is clear that one can derive an additional first integral via Eq. (\ref{thetadef}), which tells us that such systems are superintegrable. 

\begin{prop}
The two-dimensional oscillator,
\begin{equation}\label{H2d}
H =  \frac{p_1^2}{2} +  \frac{p_2^2}{2} +  V_1(q_1) + V_2(q_2),
\end{equation} where \((q_1,q_2,p_1,p_2)\) are coordinates in some region of \(\mathbb{R}^4\), \(V_1 \in C^\infty(\Gamma_1,\mathbb{R})\), \(V_2 \in C^\infty(\Gamma_2,\mathbb{R})\), with \(\Gamma_1, \Gamma_2 \subset \mathbb{R}\), is superintegrable if the quantity \(\Theta\) defined in Eq. (\ref{thetadef}) is not identically zero. 
\end{prop}

Although the above-mentioned assertion tells us that these systems are superintegrable, it does not yet describe the significance of the quantity \(\Theta\). We now prove the following result: 

\begin{theorem}\label{theorem1}
Consider the two-dimensional oscillator with Hamiltonian given by Eq. (\ref{H2d}). The first integral \(\Theta\) defined in Eq. (\ref{thetadef}) is
\begin{equation}
\Theta = \theta_1 - \theta_2,
\end{equation} where \((\theta_1,\theta_2)\) are the angle variables corresponding to the action variables \((I_1,I_2) \sim (H_1,H_2)\).
\end{theorem}

\textit{Proof -} For the system described by a Hamiltonian of the form given in Eq. (\ref{H2d}) and defined on some part of \(\mathbb{R}^4\), we introduce within the same domain, the variables
\begin{equation}
I_1 := H_1 =  \frac{p_1^2}{2} + V_1(q_1), \hspace{6mm} I_2 := H_2 = \frac{p_2^2}{2} + V_2(q_2).
\end{equation}
We then define the corresponding angles as
\begin{equation}
\theta_1 = -\frac{\partial}{\partial I_1} \int p_1 dq_1 =  -\frac{\partial}{\partial I_1} \int^{q_1} \sqrt{2I_1 - 2V_1(\zeta) } d\zeta = - \frac{1}{\sqrt{2}}\int^{q_1} \frac{d\zeta}{\sqrt{I_1 - V_1(\zeta) }} ,
\end{equation}
\begin{equation}
\theta_2 = -\frac{\partial}{\partial I_2} \int p_2 dq_2 =  -\frac{\partial}{\partial I_2} \int^{q_2} \sqrt{2I_2 - 2V_2(\zeta) } d\zeta = - \frac{1}{\sqrt{2}}\int^{q_2} \frac{d\zeta}{\sqrt{I_2 - V_2(\zeta) }},
\end{equation} with Poisson-bracket relations \(\{\theta_j,I_k\}_{\rm P.B.} = \delta_{j,k}\), \(\{I_j,I_k\}_{\rm P.B.} = \{\theta_j,\theta_k\}_{\rm P.B.} = 0\), for \(j,k = 1,2\). Next, we go on to the evaluation of \(\Theta\) as presented in Eq. (\ref{thetadef}), for which we need to first compute the quantity \(\overline{\Delta}\). In the present case, we choose \(x_1 = p_1\), \(x_2 = p_2\), \(\xi_1 := x_3 = q_1\), \(\xi_2:= x_4 = q_2\), while \(F_1 = H_1\) and \(F_2 = H_2\). This gives 
\begin{equation}
\overline{\Delta} = \frac{\partial(F_1,F_2)}{\partial(x_1,x_2)} =  \frac{\partial(H_1,H_2)}{\partial(p_1,p_2)} = p_1 p_ 2.
\end{equation}
Substituting this into Eq. (\ref{thetadef}) gives 
\begin{equation}\label{thetadef1}
\Theta = \int \frac{1}{p_1p_2} (p_1 dq_2 - p_2 dq_1),
\end{equation} which is just
\begin{eqnarray}
\Theta &=& \int \frac{dq_2}{p_2} - \int \frac{dq_1}{p_1}, \nonumber \\
&=& \frac{1}{\sqrt{2}} \int^{q_2} \frac{d\zeta}{\sqrt{I_2 - V_2(\zeta) }} - \frac{1}{\sqrt{2}} \int^{q_1} \frac{d\zeta}{\sqrt{I_1 - V_1(\zeta) }}, 
\end{eqnarray} thereby giving \(\Theta = \theta_1 - \theta_2\), completing the proof. \\

Quite remarkably, first integrals which were functions of the difference of angle variables were presented earlier in \cite{2d6,2d7,2d8,2d9,2d10,2d11}, although the explicit connection with the last multiplier was not mentioned. As a simple demonstration of the framework discussed above, let us consider the two-dimensional isotropic oscillator, with the Hamiltonian that reads as \(H = H_1 + H_2\), where
\begin{eqnarray}
 H_1=\frac{{p^2_1}+{q^2_1}}{2},\quad\quad\quad H_2=\frac{{p^2_2}+{q^2_2}}{2}.
\end{eqnarray} The superintegrability of this system is well known because it admits a \(U(2)\) symmetry, leading to conservation of the individual energies \(H_1\) and \(H_2\), together with the angular momentum \(L =  q_1 p_2 - q_2 p_1\) and the Fradkin tensor \(A = p_1p_2+q_1q_2\) \cite{ANISO,ANISO1}. In the present case, \(\Theta\) is obtained to be
\begin{eqnarray}
\Theta &=&\int^{q_2} \frac{d\zeta}{\sqrt{2I_2-\zeta^2}}- \int^{q_1}\frac{d\zeta}{\sqrt{2I_1-\zeta^2}} \nonumber \\
&=& \tan^{-1}{\left(\frac{p_2 q_1-p_1 q_2}{p_1p_2+q_1q_2}\right)}, \label{Thetalho}
\end{eqnarray} where \(I_{1,2} = H_{1,2}\). The first integral we found above is just a function of the ratio \(L/A\), where \(L\) and \(A\) are individually conserved. Further, one may verify that \(\{I_1,\Theta\}_{\rm P.B.}=-1=\{I_2,\Theta\}_{\rm P.B.}\), which implies \(\{H,\Theta\}_{\rm P.B.} = \{I_1 + I_2,\Theta\}_{\rm P.B.} = 0\), and
\begin{equation}
\{\theta_1,I_1\}_{\rm P.B.}=\{\theta_2,I_2\}_{\rm P.B.} = 1.
\end{equation}

\section{Two-dimensional oscillators with separable Hamiltonians}\label{consec}
We now apply the formalism discussed in Section (\ref{DSsec1}) to certain examples of two-dimensional unit-mass oscillators. We begin with the simple case of a linear but anisotropic oscillator on the plane. 

\subsection{Anisotropic oscillator}
The two-dimensional anisotropic oscillator is described by \(H = H_1 + H_2\), where
\begin{eqnarray}
H_1=\frac{{p^2_1}+\omega_1{q^2_1}}{2},\quad\quad\quad H_2=\frac{{p^2_2}+\omega_2{q^2_2}}{2},
\end{eqnarray} where \(\omega_1\) and \(\omega_2\) are real and positive numbers with \(\omega_1 \neq \omega_2\). The system is just two copies of the one-dimensional oscillator, but with differing frequencies. Consequently, the anisotropic oscillator is not a central-force problem; its angular momentum is not conserved. However, the system is still superintegrable because it admits a hidden \(SU(2)\) symmetry \cite{ANISO1}. We shall now demonstrate its superintegrability based on the existence of a third integral of motion.\\

Since the dynamics is Hamiltonian, the dynamical vector field has zero divergence. As a consequence, the last multiplier \(M\) is a constant, which can be set to one. Then, Eq. (\ref{thetadef}) describing the third integral of motion leads to
\begin{eqnarray}
\Theta =\frac{\tan ^{-1}\left(\frac{\sqrt{\omega_1} q_1}{p_1}\right)}{\sqrt{\omega_1}}-\frac{\tan ^{-1}\left(\frac{\sqrt{\omega_2} q_2}{p_2}\right)}{\sqrt{\omega_2}}.
\end{eqnarray}
The fact that \(\Theta\) obtained above is indeed a first integral can be independently verified from the fact that \(\{H_1+H_2,\Theta\}_{\rm P.B.}=0\), meaning \(\dot{\Theta} = 0\). Putting \(\omega_1 = \omega_2 = \omega\) (isotropic case) leads to Eq. (\ref{Thetalho}) using the familiar addition formula: \(\tan^{-1} x - \tan^{-1} y = \tan^{-1} \Big(\frac{x-y}{1 + xy}\Big)\), upon choosing \(\omega = 1\). \\

 Let us decompose \(\Theta\) into two parts as
\begin{equation}
\theta_1 = \frac{\tan ^{-1}\left(\frac{\sqrt{\omega_1} q_1}{p_1}\right)}{\sqrt{\omega_1}}, \hspace{5mm} \theta_2 = \frac{\tan ^{-1}\left(\frac{\sqrt{\omega_2} q_2}{p_2}\right)}{\sqrt{\omega_2}},
\end{equation} such that \(\Theta = \theta_1 - \theta_2\), where \(\theta_1\) is a function of \((q_1,p_1)\) and \(\theta_2\) is a function of \((q_2,p_2)\). Then, one can verify that
\begin{equation}
\{\theta_1 , H_1\}_{\rm P.B.} = \{\theta_2,H_2\}_{\rm P.B.} = 1,
\end{equation} i.e., one can perform the canonical transformations \((q_1,p_1) \mapsto (\theta_1,H_1)\) and \((q_2,p_2) \mapsto (\theta_2,H_2)\) in which \(H_1\) and \(H_2\) are conserved. But then, these are precisely the action-angle variables, meaning that \(\Theta = \theta_1 - \theta_2\) is just the difference between the two angle variables.

\subsection{Holt potential}
Next, we consider the Holt potential for which the Hamiltonian is (see for example \cite{2d11})
\begin{eqnarray}
H=\frac{1}{2}\left(p_1^2+p_2^2\right)+\frac{1}{2}\left(q_1^2+4 q_2^2\right)+\frac{\delta}{q_1^2}, \hspace{6mm} \delta > 0, 
\end{eqnarray} which is clearly separable as
\begin{equation}
H_1 = \frac{p_1^2}{2} + \frac{q_1^2}{2} + \frac{\delta}{q_1^2}, \hspace{6mm} H_2 =  \frac{p_2^2}{2} + 2 q_2^2.
\end{equation}
Then, the conserved quantity \(\Theta\) turns out to be
\begin{equation}
\Theta = \frac{i}{2}\ln \left(\frac{2 \delta}{q_1^2}-2 i q_1 p_1+p_1^2-q_1^2\right) - \frac{1}{2}\tan ^{-1}\left(\frac{2 q_2}{p_2}\right),
\end{equation} which is just \(\Theta = \theta_1 - \theta_2\), where
\begin{equation}
\theta_1 =  \frac{i}{2}\ln \left(\frac{2 \delta}{q_1^2}-2 i q_1 p_1+p_1^2-q_1^2\right), \hspace{7mm} \theta_2 = \frac{1}{2}\tan ^{-1}\left(\frac{2 q_2}{p_2}\right).
\end{equation}
These angles satisfy \(\{\theta_1,I_1\}_{\rm P.B.} = \{\theta_2, I_2\}_{\rm P.B.} = 1\), where \(I_{1,2} = H_{1,2}\). It should be noticed that the argument inside the logarithm appearing in the expression for \(\theta_1\) is complex, and one can therefore express it alternatively as \(\ln ({\rm Modulus}) + i ({\rm Phase})\). Thus, \(\theta_1\) has an imaginary part, proportional to the logarithm of the modulus of the argument of the \(\ln\) function appearing in the definition \(\theta_1\) above. It turns out that if the potential contains terms with negative powers of the coordinate variable, then the corresponding angle variable admits an imaginary part. This can also be observed from the next example. 

\subsection{Two-dimensional isotonic oscillator}
We now consider a two-dimensional generalization of the isotonic oscillator. The Hamiltonian is \(H = H_1 + H_2\), where
\begin{eqnarray}
H_1={p^2_1}+a{q^2_1}+\frac{b}{{q^2_1}},\quad\quad\quad H_2={p^2_2}+a{q^2_2}+\frac{b}{{q^2_2}},
\end{eqnarray}
where \(a, b > 0\). Once again the last multiplier is \(M = 1\); this leads to the complex-valued first integral that reads
\begin{equation}
\Theta=-\frac{i}{4b} \left(-\sqrt{b} \ln A 
   +\sqrt{b} \ln B +2 i \sqrt{b} \tan ^{-1} C-2 i
   \sqrt{b} \tan ^{-1}D \right),
\end{equation}
where 
\begin{equation}\nonumber
A = -\frac{a^2}{q_1^4}-\frac{2 p_1^2 \left(a+3 b q_1^4\right)}{q_1^2}-2 a b+8 i b^{3/2} p_1 q_1^3+3 b^2
   q_1^4-p_1^4,
\end{equation}
\begin{equation}\nonumber
B = -\frac{a^2}{q_2^4}-\frac{2 p_2^2 \left(a+3 b q_2^4\right)}{q_2^2}-2 a b+8 i b^{3/2} p_2 q_2^3+3 b^2
   q_2^4-p_2^4,
\end{equation}
\begin{equation}\nonumber
C = \frac{2 \sqrt{b} q_1^2 \left(p_1 q_1-i \sqrt{b} q_1^2\right)}{a+b q_1^4+p_1^2 q_1^2}, \hspace{6mm} D = \frac{2 \sqrt{b} q_2^2 \left(p_2 q_2-i \sqrt{b} q_2^2\right)}{a+b q_2^4+p_2^2 q_2^2}.
\end{equation}
We have \(\{H_1+H_2,\Theta\}_{\rm P.B.}=0\). Now, if we define 
\begin{equation}
\theta_1 =-\frac{i}{4b} \left(-\sqrt{b} \ln A 
    +2 i \sqrt{b} \tan ^{-1} C\right), \hspace{6mm} \theta_2 =-\frac{i}{4b} \left(-\sqrt{b} \ln B 
    +2 i \sqrt{b} \tan ^{-1} D\right),
\end{equation} then one has \(\Theta = \theta_1 - \theta_2\) and \(\{\theta_1,I_1\}_{\rm P.B.} = \{\theta_2,I_2\}_{\rm P.B.} = 1\), where \(I_1 = H_1/2\) and \(I_2 = H_2/2\). The pairs of variables \((\theta_1,I_1)\) and \((\theta_2,I_2)\) should be viewed as (complex) action-angle variables for the two one-dimensional isotonic oscillators.

\subsection{A two-dimensional nonlinear system}\label{nonlinsec}
Let us consider a nonlinear two-dimensional system with Hamiltonian \(H = H_1 + H_2\), where \cite{2d0,2d10}
\begin{eqnarray}
H_1=p_1^2+q_1^\alpha,\quad\quad\quad H_2= p_2^2+q_2^\beta,
\end{eqnarray} where \(\alpha\) and \(\beta\) are some rational numbers. Clearly, the Hamiltonian is well defined in a certain region of \(T^*\mathbb{R}^2\).\\

 Due to energy conservation in Hamiltonian dynamics, \(H_1\) and \(H_2\) are both conserved. We shall now determine the third conserved quantity using the Jacobi last multiplier. Once again, the last multiplier is \(M=1\). In order to determine the first integral \(\Theta\) [Eq. (\ref{thetadef})], we identify \(x_1 = p_1\), \(x_2 = p_2\), \(\xi_1 := x_3 = q_1\), and \(\xi_2 := x_4 = q_2\). Then we have \(\overline{\Delta} = 4 p_1 p_2\), and using this in Eq. (\ref{thetadef}), we find
\begin{eqnarray}
\Theta =\frac{1}{4} \bigg( \int^{q_2}\frac{d\zeta}{\sqrt{H_2-\zeta^\beta}}-\int^{q_1}\frac{d\zeta}{\sqrt{H_1-\zeta^\alpha}} \bigg),
\end{eqnarray} which can be immediately identified to be of the form \(\Theta = \theta_1 - \theta_2\), where
\begin{equation}
\theta_1 = -\frac{1}{4} \int^{q_1} \frac{d\zeta}{\sqrt{H_1-\zeta^\alpha}}, \hspace{6mm} \theta_2 = -\frac{1}{4} \int^{q_2} \frac{d\zeta}{\sqrt{H_2-\zeta^\beta}},
\end{equation} consistent with our previous assertion. Integrals of the above-mentioned type can be evaluated in terms of elementary functions using the Chebyshev's theorem on differential binomials \cite{cheby}, as has been considered in \cite{2d10,2d11}. We have verified that it is possible to express the variables \(\theta_1\) and \(\theta_2\) as
 \begin{equation}\label{nonlinangle}
 \theta_1 = \frac{p_1 q_1 \times {_2F_1}\left(1,\frac{1}{2}+\frac{1}{\alpha};1+\frac{1}{\alpha};\frac{1}{p_1^2 q_1^{-\alpha}+1}\right)}{q_1^\alpha+p_1^2}, \hspace{6mm} \theta_2 = \frac{p_2 q_2 \times
   {_2F_1}\left(1,\frac{1}{2}+\frac{1}{\beta};1+\frac{1}{\beta};\frac{1}{p_2^2 q_2^{-\beta}+1}\right)}{q_2^\beta+p_2^2},
 \end{equation}
where $_2F_1(a,b;c;z)$ is the hypergeometric function. It may also be verified that \(\{H_1+H_2,\Theta\}_{\rm P.B.}=0\), demonstrating that \(\Theta\) is indeed a first integral. 

\subsection{Purely-nonlinear oscillators on the plane}
We consider a generalization of the model presented in the previous subsection. The Hamiltonian is \(H = H_1 + H_2\), where
 \begin{eqnarray}
H_1=p_1^2+|q_1|^\alpha,\quad\quad\quad H_2= p_2^2+|q_2|^\beta,
\end{eqnarray} where \(\alpha\) and \(\beta\) are rational numbers greater than two. Its one-dimensional counterpart was studied in the context of action-angle variables in \cite{Ghosh}. This particular model, unlike the one discussed in Section (\ref{nonlinsec}), is defined everywhere on \(T^*\mathbb{R}^2\), owing to the fact that \(|q_j|\) for \(j \in \{1,2\}\) only takes the absolute value, and that \(\alpha, \beta > 2\). It should be emphasized that due to the modulus sign in the potential functions, they are not of the \(C^\infty\) class, unless \(\alpha\) and \(\beta\) are even numbers. However, the result of Theorem (\ref{theorem1}) still holds.\\

Let us find the conserved quantity \(\Theta\) directly by introducing action-angle variables as \(\Theta = \theta_1 - \theta_2\), where
\begin{equation}
\theta_1 = - \int^{q_1} \frac{d\zeta}{\sqrt{2I_1-|\zeta|^\alpha}}  \hspace{7mm} \theta_2 = -  \int^{q_2} \frac{d\zeta}{\sqrt{2I_2-|\zeta|^\beta}},
\end{equation} with \(I_{1,2} = H_{1,2}/2\). The above-mentioned integrals can be expressed in terms of generalized trigonometric functions \cite{Ghosh,GTF,GTF1}. For simplicity, we choose the initial conditions as \(q_1(0) = a\), \(q_2(0) = b\), \(p_1(0) = 0\), \(p_2(0) = 0\), where \(a, b > 0\). Then, \(I_1 = |a|^\alpha/2 \) and \(I_2 = |b|^\beta/2 \), which gives \cite{Ghosh}
\begin{equation}\label{Thetagensine}
\Theta = b^{1 - \beta/2} \sin^{-1}_{2,\beta} \bigg(\frac{q_2}{b}\bigg) - a^{1 - \alpha/2} \sin^{-1}_{2,\alpha} \bigg(\frac{q_1}{a}\bigg).
\end{equation}
Here, we have defined the generalized sine function \(\sin_{2,r} (\cdot)\) as the inversion of
\begin{equation}
\sin^{-1}_{2,r} (x) := \int_0^x \frac{d\zeta}{\sqrt{1 - |\zeta|^r}}, \hspace{6mm} r > 1. 
\end{equation}

\subsection{P\"oschl–Teller potential in two dimensions}
Let us consider a two-dimensional oscillator with Hamiltonian of the form \(H = H_1 + H_2\), where
\begin{equation}
H_1 = \frac{p_1^2}{2}-\omega_1\;\textrm{sech}^2 q_1, \quad\quad\quad H_2 = \frac{p_2^2}{2}-\omega_2\;\textrm{sech}^2 q_2.
\end{equation}
Here, \(\omega_{1,2} = \frac{\lambda_{1,2}(\lambda_{1,2} + 1)}{2}\), where \(\lambda_{1,2}\) are real parameters, typically positive integers. Employing the last multiplier, we find the following integral of motion: 
\begin{equation}
\Theta = \frac{1}{\sqrt{p_2^2-2\omega_2\; \textrm{sech}^2 q_2}}\textrm{arccoth}\left(\frac{p_2\;\textrm{coth} q_2}{\sqrt{p_2^2-2\omega_2\; \textrm{sech}^2 q_2}}\right) - \frac{1}{\sqrt{p_1^2-2\omega_1\; \textrm{sech}^2q_1}}\textrm{arccoth}\left(\frac{p_1\;\textrm{coth}q_1}{\sqrt{p_1^2-2\omega_1\; \textrm{sech}^2q_1}}\right).
\end{equation}
This can be expressed as \(\Theta = \theta_1 - \theta_2\), with
\begin{eqnarray}
    \theta_{1,2}=-\frac{1}{\sqrt{2}}\int^{q_{1,2}} \frac{d\zeta}{\sqrt{I_{1,2}-V_{1,2}(\zeta)}},
\end{eqnarray} where \(I_{1,2} = H_{1,2}\) and $V_{1,2}(\zeta)=-\omega_{1,2}\; \textrm{sech}^2 \zeta $. Thus, we finally have
\begin{equation}
    \theta_{1,2}=-\frac{1}{\sqrt{p_{1,2}^2-2\omega_{1,2}\; \textrm{sech}^2q_{1,2}}}\textrm{arccoth}\left(\frac{p_{1,2}\;\textrm{coth}q_{1,2}}{\sqrt{p_{1,2}^2-2\omega_{1,2}\; \textrm{sech}^2q_{1,2}}}\right).
\end{equation}

\subsection{Two pendulum potentials}
We consider a two-dimensional system with a Hamiltonian of the form \(H = H_1 + H_2\), where
\begin{equation}
H_1=\frac{p_1^2}{2}+k_1\left(1-\cos q_1\right), \quad\quad\quad H_2=\frac{p_2^2}{2}+k_2\left(1-\cos q_2\right), \quad\quad\quad k_1,k_2 > 0.
\end{equation}
The additional conserved quantity comes out to be \(\Theta = \theta_1 - \theta_2\), with
\begin{eqnarray}
    \theta_{1,2}=-\frac{1}{\sqrt{2}}\int^{q_{1,2}} \frac{d\zeta}{\sqrt{I_{1,2}-V_{1,2}(\zeta)}},
\end{eqnarray} where \(I_{1,2} = H_{1,2}\) and \(V_{1,2}(\zeta) = k_{1,2}\left(1-\cos \zeta \right)\). This gives
\begin{equation}
\Theta = \sqrt{\frac{2}{I_2}}\mathcal{F}\left(\frac{q_2}{2},\frac{2 k_2}{I_2}\right)  - \sqrt{\frac{2}{I_1}}\mathcal{F}\left(\frac{q_1}{2},\frac{2 k_1}{I_1}\right),
\end{equation}
where $\mathcal{F}$ is the incomplete elliptic function of the first kind.

\section{Position-dependent-mass oscillators}\label{PDMsec}
Let us now consider position-dependent-mass oscillators in two dimensions \cite{2d11}. Although one may still encounter Hamiltonians that are separable, i.e., are of the form \(H = H_1 + H_2\), the structure is not exactly of the form given in Eq. (\ref{H2d}) due to the presence of position-dependent masses. As we show via an example, it may still be possible to express the first integral \(\Theta\) as the difference between the two angles. Consider the case where the Hamiltonian reads
\begin{equation}\label{L1}
H=\frac{p_1^2}{2m_1(q_1)}+\frac{p_2^2}{2m_2(q_2)}+\frac{1}{2}m_1(q_1)q_1^2+
\frac{1}{2}m_2(q_2)q_2^2,
\end{equation}
where $(q_j, p_j)$ for $j\in \{1,2\}$ satisfy $\{q_j, p_k\}_{\rm P.B.}=\delta_{j,k}$ and
$\{q_j,q_k\}_{\rm P.B.}=\{p_j, p_k\}_{\rm P.B.}=0$. The Hamilton's equations are then given by 
\begin{equation}\label{L2}
\dot{q}_j=\frac{p_j}{m_j(q_j)}, \hspace{6mm} \dot{p}_j=\frac{m_j^\prime(q_j)}{2m_j(q_j)^2}p_j^2-\frac{1}{2} \frac{d}{dq_j}\big(m_j(q_j) q_j^2\big),\hspace{6mm} j\in \{1,2\}.
\end{equation}
Suppose the mass functions are such that
\begin{equation}\label{C6}
\left[1+\frac{q_j}{2}\frac{m_j^\prime(q_j)}{m_j(q_j)}\right]=\lambda_j, \hspace{6mm} j\in \{1,2\},
\end{equation}
where $\lambda_j$ are constants. Following \cite{2d11}, for the purpose of illustration we take $\lambda_1=2$ and $\lambda_2=-1$, along with (integration constants) $m_{01} = m_{02}=1$; this corresponds to the Hamiltonian that reads
\begin{equation}\label{H1111}
H=\frac{p_1^2}{2q_1^2}+\frac{p_2^2}{2q_2^{-4}}+\frac{q_1^4}{2}+\frac{2}{q_2^2}.\end{equation} 

We now find the additional first integral \(\Theta\). For that, we identify \(x_1 = p_1\), \(x_2 = p_2\),  \(\xi_1 := x_3 = q_1\),  \(\xi_2 := x_4 = q_2\), \(F_1 = H_1\), and \(F_2 = H_2\). This gives 
\begin{eqnarray}
\overline{\Delta}=\frac{\partial(F_1,F_2)}{\partial(x_1,x_2)}=\frac{p_1p_2 q_2^4}{q_1^2}, \hspace{6mm} \overline{X}_1 = \frac{p_1}{q_1^2}, \hspace{6mm}   \overline{X}_2 = p_2 q_2^4.
\end{eqnarray}
Putting these into Eq. (\ref{thetadef}), after some straightforward manipulations we get
\begin{eqnarray}
\Theta= - \frac{1}{2} \tan^{-1} \bigg(\frac{p_1}{q_1^3}\bigg) - \frac{1}{2}  \tan^{-1} \bigg(\frac{p_2 q_2^3}{2}\bigg).
\end{eqnarray} This lets us identify
\begin{equation}
\theta_1 = - \frac{1}{2} \tan^{-1} \bigg(\frac{p_1}{q_1^3}\bigg), \hspace{6mm} \theta_2 = \frac{1}{2}  \tan^{-1} \bigg(\frac{p_2 q_2^3}{2}\bigg).
\end{equation} These angle variables satisfy \(\{\theta_1,I_1\}_{\rm P.B.} = \{\theta_2,I_2\}_{\rm P.B.} = 1\) for \(I_{1,2} = H_{1,2}\). Various other examples can be worked out and Eq. (\ref{thetadef}) gives an additional first integral as long as one has two functionally-independent first integrals \(H_1\) and \(H_2\) to begin with.

\section{Superintegrability of the Bateman pair}\label{Bsec}
We now demonstrate that the Bateman pair is superintegrable. Let us recall that the Bateman pair consists of the following linear and uncoupled systems \cite{bateman}:
\begin{equation}\label{batemaneom2}
\ddot{x} + \gamma \dot{x} + \omega^2 x = 0, \hspace{6mm} \ddot{y} - \gamma \dot{y} + \omega^2 y = 0,
\end{equation} where the \(x\)-oscillator is found to dissipate its mechanical energy with damping constant \(\gamma > 0\), while the \(y\)-oscillator gains energy with gain constant \(\gamma\). In other words, one may think of the Bateman pair to be a pair of linear oscillators where one of them gains the energy that is lost by the other. \\

Consider the Hamiltonian \cite{bateman1}
\begin{eqnarray}
H_{\rm B}=p_x p_y+\omega^2 x y-\gamma (x p_x-y p_y).
\end{eqnarray}
The Hamilton's equations give 
\begin{eqnarray}
\dot{x} &=& \frac{\partial H_{\rm B}}{\partial p_x} = p_y - \gamma x, \hspace{7mm} \dot{p}_x = -\frac{\partial H_{\rm B}}{\partial x} = - \omega^2 y + \gamma p_x, \\
\dot{y} &=& \frac{\partial H_{\rm B}}{\partial p_y} = p_x + \gamma y, \hspace{7mm} \dot{p}_y = -\frac{\partial H_{\rm B}}{\partial y} = - \omega^2 x - \gamma p_y,
\end{eqnarray} from which Eq. (\ref{batemaneom2}) follows straightforwardly. Therefore, \(H_{\rm B}\) shall be called the Bateman Hamiltonian. Now, although the individual mechanical energies are not conserved, one finds the following two conserved quantities by inspection:
\begin{equation}
H_\omega = p_x p_y+\omega^2 x y, \hspace{7mm} H_\gamma = -\gamma (x p_x-y p_y),
\end{equation} such that \(H_{\rm B}  = H_\omega + H_\gamma\). Here, $H_\omega$ and $H_\gamma$ are functionally independent and individually conserved, i.e.,
\begin{eqnarray}
\{H_\omega,H_{\rm B}\}_{\rm P.B.}=0=\{H_\gamma,H_{\rm B}\}_{\rm P.B.}.
\end{eqnarray}
We shall use the knowledge of these conserved quantities to obtain other conserved quantities with the aid of the last multiplier. One should note that unlike the examples considered so far, the Bateman pair is not described by a Hamiltonian of the form \(H = H_x + H_y\), where \(H_x\) and \(H_y\) depend respectively on the variables \((x,p_x)\) and \((y,p_y)\) only. As a consequence of this, the conserved quantity that we shall now obtain is not of the form dictated by \(\Theta = \theta_1 - \theta_2\). \\

To apply the formalism discussed in Section (\ref{DSsec}), we perform the identifications that go as
\begin{eqnarray}
x_1 = x,\quad x_2 = p_y,\quad x_3 = y,\quad x_4 = p_x,\quad F_1 = H_\omega,\quad F_2 = H_\gamma.
\end{eqnarray}   
 Now it should be emphasized that although the dynamics of the two oscillators is not individually divergence free, the combined dynamics described by \(H_{\rm B}\) has zero divergence, because it is of Hamiltonian nature. More specifically, the divergence coming from the \(x\)-oscillator is \(-\gamma\), while that coming from the \(y\)-oscillator is \(+\gamma\); these exactly cancel each other to give zero (overall) divergence. Consequently, we get \(M = 1\). We further have
\begin{eqnarray}
\overline{\Delta}=\frac{\partial(F_1,F_2)}{\partial(x_1,x_2)}=\gamma\left(p_x^2+\omega^2 y^2\right).
\end{eqnarray}
This gives a conserved quantity that reads
\begin{eqnarray}
\Theta_1&=&\int \frac{1}{\gamma\left(p_x^2+\omega^2 y^2\right)}\left[\left(p_x+\gamma y\right)dp_x-\left(\gamma p_x-\omega^2 y\right)dy\right] \nonumber \\
&=&\int \frac{1}{\gamma\left(p_x^2+\omega^2 y^2\right)}\left[\left(p_xdp_x+\omega^2 y dy\right)+\gamma\left(y dp_x-p_x dy\right)\right] \nonumber \\
&=&\frac{1}{2\gamma}\ln\left(p_x^2+\omega^2 y^2\right)+\frac{1}{\omega}\tan^{-1}\bigg(\frac{p_x}{\omega y}\bigg),
\end{eqnarray} and proves the superintegrability of the Bateman pair. Interestingly, one can find another conserved quantity which reads
\begin{eqnarray}
\Theta_2=\frac{1}{2\gamma}\ln\left(p_y^2+\omega^2 x^2\right)-\frac{1}{\omega}\tan^{-1}\bigg(\frac{p_y}{\omega x}\bigg),
\end{eqnarray} and satisfies
\begin{eqnarray}
\{\Theta_1,\Theta_2\}_{\rm P.B.}=\frac{(\omega^2-\gamma^2)H_\gamma-2\gamma^2H_\omega}{\gamma\left(\gamma^2 H_\omega^2+\omega^2 H_\gamma^2\right)}.
\end{eqnarray}
Obviously, all the conserved quantities, i.e., \(\{H_\omega, H_\gamma, \Theta_1, \Theta_2\}\) are not independent of each other.

\section{Discussion}\label{dsec}
In this paper, we explored the role played by the last multiplier in the context of two-dimensional superintegrable systems. For most part, we considered systems which can be described by a Hamiltonian of the form given in Eq. (\ref{H2d}), i.e., \(H = H_1 + H_2\), where \(H_{1,2}\) represent the Hamiltonians of two independent unit-mass oscillators in one dimension. Using the last multiplier, one can associate an additional first integral \(\Theta\) to such systems, apart from the usual ones which are \(H_1\) and \(H_2\). We worked out various examples, and also showed that the additional first integral bears an interesting interpretation: it is the difference between the angles \(\theta_1\) and \(\theta_2\), associated with action variables \(I_1 \sim H_1\) and \(I_2 \sim H_2\), respectively. We then considered two-dimensional position-dependent-mass oscillators and demonstrated the superintegrability of a simple example. Finally, we considered the Bateman pair which is a very different kind of system owing to the fact that the individual oscillators undergo dissipative dynamics and the Hamiltonian is not separable. However, we demonstrated that the formalism involving the last multiplier successfully leads to the emergence of additional first integrals, indicating that the Bateman pair is a (maximally) superintegrable system. \\

It is curious to ask if the dynamics of the systems considered in this work can be described in the framework of Nambu mechanics \cite{Nambu1,Nambu2,Nambu3}. In fact, Nambu mechanics turns out to be useful particularly in the case of integrable systems, where different conserved quantities can serve the role of `Nambu-Hamiltonians' and are set on the same footing as the usual Hamiltonian. Since the phase space at present has a Poisson structure endowed to it due to the symplectic structure, we may also (at least locally) associate a Nambu bracket with it. Further, because the phase spaces of all systems considered in this paper are four-dimensional, one would naturally define a Nambu 4-bracket on the algebra of functions. Consider for instance, the Bateman pair which was studied in Section (\ref{Bsec}). If we define the quantity
\begin{eqnarray}
\overline{H}= H^2_\omega + \left( \frac{\omega}{\gamma}\right)^2 H_\gamma^2,
\end{eqnarray}
then the Nambu bracket \(\{\cdot,\cdot,\cdot,\cdot\}_{\rm N.B.}:C^\infty(\mathbb{R}^4,\mathbb{R})^{\times 4} \rightarrow C^\infty(\mathbb{R}^4,\mathbb{R})\) is defined as 
\begin{eqnarray}\label{Nambu}
\{f_1 ,f_2 ,f_3,f_4\}_{\rm N.B.}= \bigg(\frac{\gamma \overline{H}}{H_{\rm B}}\bigg) \frac{\partial(f_1 ,f_2 ,f_3,f_4)}{\partial(x_1 ,x_2 ,x_3,x_4)},
\end{eqnarray} for \(f_1, f_2, f_3, f_4 \in C^\infty(\mathbb{R}^4,\mathbb{R})\). Here, the subscript `N.B.' denotes Nambu bracket. One can easily check that for any observable \(f \in C^\infty(\mathbb{R}^4,\mathbb{R})\), its evolution under the Bateman dynamics is
\begin{eqnarray}\label{Nambu1}
\frac{df}{dt}=\{f,H_\gamma,\Theta_1,\Theta_2\}_{\rm N.B.}. 
\end{eqnarray}
Generalizations to other examples is straightforward. It may be remarked that a Nambu bracket may involve subordinate Poisson brackets \cite{guha}. \\

\textbf{Acknowledgements:} We thank Jasleen Kaur for carefully reading the manuscript. A.S. would like to acknowledge the financial support from IIT Bhubaneswar, in the form of an Institute Research Fellowship. A.G. would like to thank Chandrasekhar Bhamidipati for many discussions during their collaboration on \cite{Ghosh} and Debashis Ghoshal for some useful remarks on purely-nonlinear oscillators. The work of A.G. is supported by the Ministry of Education (MoE), Government of India, in the form of a Prime Minister's Research Fellowship (ID: 1200454).  \\

\textbf{Data availability statement:} We do not analyze or generate any datasets, because our work proceeds within a theoretical and mathematical approach.\\

\textbf{Conflicts of interest:} The authors certify that they have no affiliations with or involvement in any
organization or entity with any financial or non-financial interest in
the subject matter or materials discussed in this manuscript.

\end{document}